\newcommand{\Msun}{\mathrm{M}_{\odot}}
\newcommand{\yr}{\mathrm{yr}}
\newcommand{\Mpc}{\mathrm{Mpc}}
\documentclass[twocolumn]{aastex63}

\received{}
\revised{}
\accepted{}
\submitjournal{ApJ}

\shorttitle{}
\shortauthors{Deme et al.}
\graphicspath{{./}{figures/}}

\usepackage{amsmath}

\begin{document}

\title{Detecting Kozai-Lidov imprints on the gravitational waves of intermediate-mass black holes in galactic nuclei}

\correspondingauthor{Barnab{\'a}s Deme}
\email{deme.barnabas@gmail.com}

\author[0000-0003-4016-9778]{Barnab{\'a}s Deme}
\affiliation{Institute of Physics, E\"otv\"os University, P\'azm\'any P. s. 1/A, Budapest, 1117, Hungary}

\author[0000-0003-0992-0033]{Bao-Minh Hoang}
\affiliation{Department of Physics and Astronomy, University of California, Los Angeles, CA90095, USA}
\affiliation{Mani L. Bhaumik Institute for Theoretical Physics, Department of Physics and Astronomy, UCLA, Los Angeles, CA 90095, USA
}

\author[0000-0002-9802-9279]{Smadar Naoz}
\affiliation{Department of Physics and Astronomy, University of California, Los Angeles, CA90095, USA}
\affiliation{Mani L. Bhaumik Institute for Theoretical Physics, Department of Physics and Astronomy, UCLA, Los Angeles, CA 90095, USA
}

\author[0000-0002-4865-7517]{Bence Kocsis}
\affiliation{Rudolf Peierls Centre for Theoretical Physics, Clarendon Laboratory, Parks Road, Oxford OX1 3PU, UK}

\begin{abstract}

A third object in the vicinity of a binary system causes variations in the eccentricity and the inclination of the binary through the Kozai-Lidov effect. We examine if such variations leave a detectable imprint on the gravitational waves of a binary consisting of intermediate mass black holes and stellar mass objects. As a proof of concept, we present an example where LISA may detect the Kozai-Lidov modulated gravitational wave signals of such sources from at least a distance of 1~Mpc if the perturbation is caused by a supermassive black hole tertiary.
Although the quick pericenter precession induced by general relativity significantly reduces the appropriate parameter space for this effect by quenching the Kozai-Lidov oscillations, we still find reasonable parameters where the Kozai-Lidov effect may be detected with high signal-to-noise ratios.

\end{abstract}


\section{Introduction}\label{intro}
According to the current paradigm, nearly all galaxies, including our own, host a supermassive black hole (SMBH) at their centers \citep{kormendy,Ghez+08,Genzel+10}.  Being the engine of galactic nuclear activity, they have a large influence both on their immediate environment  \citep[e.g.,][]{inoue} and on more extended scales which leads to correlations between the SMBH and the host galaxy properties \citep{king}. Observations such as periodic AGN variability show that some SMBHs are found in binaries \citep{kelley}. They are the natural consequences of galaxy mergers predicted by the $\Lambda$CDM model \citep[e.g.,][]{DiMatteo+05,Hopkins+06,Robertson+06}. The SMBH at the center of the Milky Way may also have a massive binary companion \citep[see][for observational constraints]{Gualandris_Merritt2009,Gualandris+2010,Naoz+20}. SMBH binaries play a key role in galaxy evolution \citep[e.g.,][]{begelman,Blecha+08}. They explain the mass deficit of stars observed in the centers of galaxies \citep{Merritt2006,Gualandris_Merritt2012}, they lead to the ejection of hyper velocity stars \citep[e.g.,][]{Yu+03,Luna+19,Rasskazov+19,Fragione+19}, affect tidal disruption and GW events \citep[e.g.,][]{Ivanov+05,Chen+09,Chen+11,Chen+13,Wegg+11,Sesana+11,Li+15,Meiron+13,Fragione+20}, and lead to an electromagnetic signature from dark matter annihilation \citep{NaozSilk14,Naoz+19}.
The inspiral of such SMBH binaries will be targets for the future space-borne gravitational wave (GW) observatory LISA\footnote{https://lisa.nasa.gov/} \citep[e.g.,][]{amaro-seoane}. 

The mass spectrum of SMBHs arguably extends down to the regime of intermediate-mass black holes (IMBHs) (see \citealt{greene} and \citealt{mezcua2017} for recent reviews). An SMBH-IMBH binary may reside in the nucleus of some galaxies. The IMBHs may form in SMBH accretion disks  \citep{Goodman_Tan2004,McKernan2012} or they may be transported to the galactic center region by infalling globular clusters that also help to form the nuclear star clusters around SMBHs \citep{pzwart2006,MBattisti2014}. 
Gravitational wave (GW) astronomy, which has recently opened a new window on the Universe \citep{abbott2016}, may directly test the existence of IMBHs in galactic nuclei. 

There are multiple dynamical processes in the nuclear regions of galaxies which may affect the binaries' GWs.
The pertubations associated with the SMBH in nuclear star clusters may be significant. The
SMBH perturber leads to the acceleration of a binary's center of mass which may be detected by LISA \citep{Yunes_Miller2011}. Furthermore, variations caused by relativistic beaming, Doppler, and gravitational redshift associated with the SMBH companion may also lead to potentially detectable signatures \citep{Meiron+2017}. 
In this paper, we examine if the Kozai-Lidov (KL) effect of the SMBH leads to detectable variations on binaries in nuclear star clusters. 

The KL mechanism has been long recognized to be one of the important dynamical processes in galactic nuclei (see \citealt{naoz2016} for a review). It describes the long-term dynamics of a hierarchical triple system in separation, i.e. when two of the bodies constitute a tight inner binary, which is orbited by a more distant tertiary (outer binary). This third object perturbs the inner binary in such a way that it exhibits eccentricity and inclination oscillations with nearly constant semi-major axis \citep{kozai1962,lidov1962}. It can be shown that the octupole-order perturbation by the third body can pump up the eccentricity to very high values close to unity \citep{lithwick2011} which leads to gravitational wave bursts during close periapsis encounters \citep{oleary,Kocsis_Levin2012}.   
The KL torque from a SMBH may result in the merger of compact object binaries (e.g. \citealt{antonini2014,hoang2018}).\footnote{The mergers in such a scenario can be further facilitated by other dynamical effects, including mass-segregation \citep{oleary}, vector resonant relaxation \citep{hamers2018}, "gas capture" binary formation in AGN disks \citep{tagawa2019} or resonant(-like) general relativistic effects \citep{naoz2013,liu2020,yun2020}.} The eccentricity oscillations from KL effects are directly detectable in the inspiral phase long before the merger of the inner binary, which causes a periodic shift in the GW strain signal \citep{hoang2019,randall2019,gupta2019}.

\begin{figure}
\includegraphics[scale=0.3]{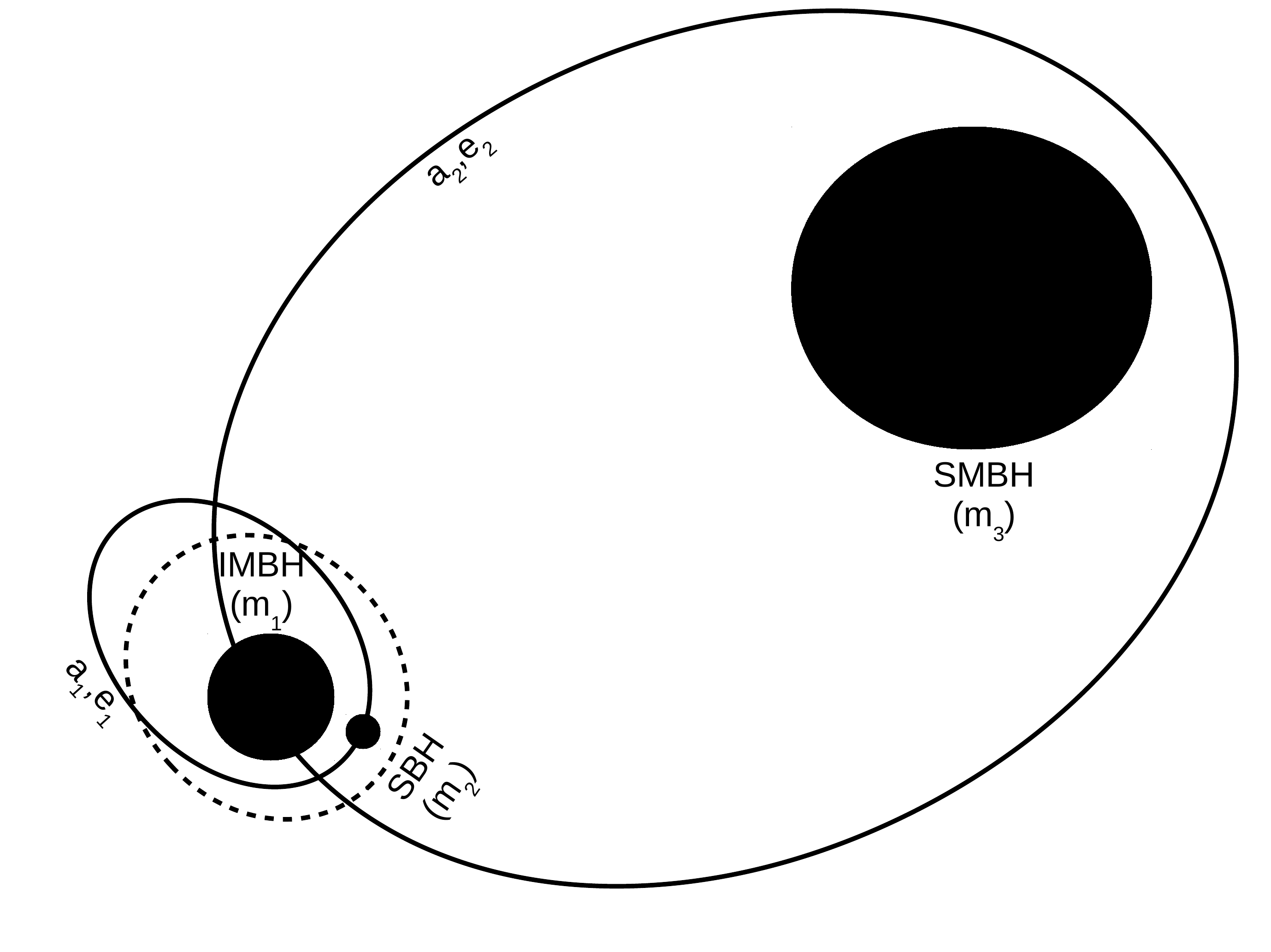}
\caption{The configuration under consideration. The dotted ellipse shows KL-modulated inner orbit.}
\label{config}
\end{figure}

In this paper we examine how the mass and initial orbital parameters of the inner binary affect LISA's ability to identify the KL effect of the SMBH on binaries in galactic nuclei (see also \cite{emami2019}). We show that if the inner binary consists of an IMBH and a stellar-mass black hole (see Fig. \ref{config}), LISA may directly detect the KL oscillations from a distance of 1 Mpc. 

This paper is structured as follows. In Section \ref{timescales} we introduce the timescales which have key role in the dynamics we investigate. In Section \ref{snr} we calculate the signal-to-noise ratio and in Section \ref{discussion} we discuss our results.

\section{Timescales and constraints}\label{timescales}
The relevant timescales for our study are the KL time ($T_\mathrm{KL}$), the general relativistic (GR) apsidal precession time ($T_\mathrm{GR}$), and the GW inspiral time ($T_\mathrm{GW}$) \citep{naoz2016,Peters1964}:
\begin{align}\label{LK}
T_\mathrm{KL}&=\frac{a_2^3(1-e_2^2)^{3/2}(m_1+m_2)^{1/2}}{G^{1/2}a_1^{3/2}m_3},\\
\label{GR1}
T_\mathrm{GR,1} &=\frac{a_1^{5/2}c^2(1-e_1^2)}{G^{3/2}(m_1+m_2)^{3/2}},   \\
\label{GW1}
T_\mathrm{GW,1} &=\frac{5c^5a_1^4}{64G^3m_1m_2(m_1+m_2)F(e_1)},\\
\label{GW2}
T_\mathrm{GW,2}&=\frac{5c^5a_2^4}{64G^3(m_1+m_2)m_3(m_1+m_2+m_3)F(e_2)},
\end{align}
where the $1$ and $2$ subscripts refer to the inner and outer binaries, respectively, and
\begin{equation}
    F(e)=\frac{1+\frac{73}{24}e^2+\frac{37}{96}e^4}{(1-e^2)^{7/2}}
\end{equation}

Long-lived triples must satisfy the Hill stability criterion, i.e.
\begin{equation}\label{Hill}
\frac{a_1}{a_2} \lesssim \frac{1-e_2}{1+e_1}\left(\frac{m_1+m_2}{3m_3}\right)^{1/3}\,,
\end{equation}
and we restrict attention to sufficiently hierarchical configurations so that we can neglect the terms beyond octupole in the expansion of the Hamiltonian
\begin{equation}\label{Hierarchical}
\frac{a_1}{a_2}\lesssim0.1 \frac{1-e_2^2}{e_2}\,.
\end{equation}
Eqs. \eqref{Hill} and \eqref{Hierarchical} show that Eq. \eqref{Hill} is always more strict if $(m_1+m_2)/m_3\lesssim0.024$. In particular, Hill-stable inner binaries with $10$ and $10^5$ $\Msun$ around a $10^8\mbox{ }\Msun$ SMBH  have $(m_1+m_2)/m_3\sim 0.001$ and automatically satisfy the hierarchy criterion.   
 
\begin{figure*}
\includegraphics[scale=1.3]{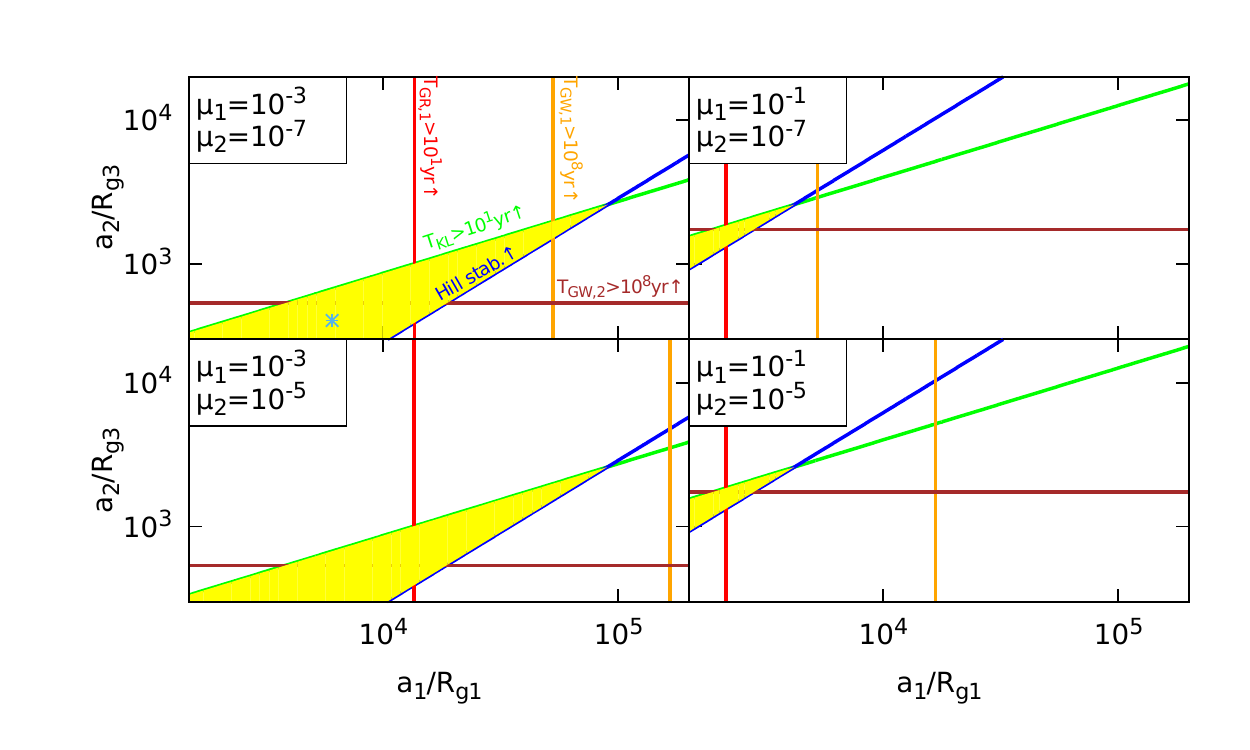}
\caption{Timescales and Hill stability criterion for repressive systems. We consider three black holes in a hierarchical configuration, where $m_3=10^8\mbox{ }\Msun$ and the masses of the other two are given with ratios $\mu_1=m_1/m_3$ and $\mu_2=m_2/m_3$ in the legends. The gravitational radius of the the tertiary, the length unit, is $R_\mathrm{g3}=0.987$ AU, while the eccentricities are set to $e_1=0.985$ and $e_2=10^{-6}$ initially. The yellow zone indicates the ideal region for detecting SMBH-induced KL oscillations, where the KL timescale is on the order of LISA's lifetime and the system is Hill-stable. The blue asterisk indicates the initial condition of a representative simulation shown in Fig. \ref{e1} ($a_1/R_\mathrm{g1}=6077.6$, $a_2/R_\mathrm{g3}=405.22$).
}
\label{heatmap}
\end{figure*}

It is useful to express Eqs. \eqref{LK}-\eqref{GW2} with distances normalized by the corresponding gravitational radii:
\begin{equation}
R_\mathrm{g1}=\frac{G(m_1+m_2)}{c^2}\,,\quad
R_\mathrm{g3}=\frac{Gm_3}{c^2}\,,\quad
\end{equation}
and with the mass ratios $\mu_1= m_1/m_3$ and $\mu_2= m_2/m_3$ as
\begin{align}
\frac{T_\mathrm{KL}}{R_\mathrm{g3}/c} &=  \frac{(1-e_2^2)^{3/2}}{(\mu_1+\mu_2)} \frac{(a_2/R_\mathrm{g3})^{3}}{(a_1/R_\mathrm{g1})^{3/2}}\,,\\
\frac{T_\mathrm{GR,1}}{R_\mathrm{g3}/c} &= (1-e_1^2) (\mu_1+\mu_2) (a_1/R_\mathrm{g1})^{5/2}\,,\\
\frac{T_\mathrm{GW,1}}{R_\mathrm{g3}/c} &= \frac{5}{64F(e_1)}
\frac{(\mu_1+\mu_2)^3}{\mu_1\mu_2}\left(\frac{a_1}{R_\mathrm{g1}}\right)^4\,,\\
\frac{T_\mathrm{GW,2}}{R_\mathrm{g3}/c} &= \frac{5}{64F(e_2)}  \frac{(a_2/R_\mathrm{g3})^4}{(\mu_1+\mu_2)(1+\mu_1+\mu_2)}\,.
\end{align}
and the Hill stability criterion reads
\begin{equation}
\frac{a_1/R_\mathrm{g1}}{a_2/R_\mathrm{g3}}<3^{-1/3}\frac{1-e_2}{1+e_1}(\mu_1+\mu_2)^{-2/3}\,.
\end{equation}

Fig. \ref{heatmap} shows the parameter space for different separations and mass ratios where these timescales are in the suitable range for the KL effect to play a role during LISA observations. We show cases where $T_\mathrm{KL}< 10 \yr$, $T_\mathrm{GR,1}> 10 \yr$,
$T_\mathrm{GW,1} > 10^8\,\yr$, and $T_\mathrm{GW,2}>10^8\,\yr$ for $m_3=10^8\mbox{ }\Msun$. The values of the timescales are chosen arbitrarily, but in the case of GR and KL times (10 years) we took into account the operational time of LISA. We also note that once $T_\mathrm{GR,1}<T_\mathrm{KL}$, the KL mechanism is quenched, i. e. the amplitude of the eccentricity oscillations is significantly damped, however, we will show that they are still detectable. The right panels show higher $\mu_1$ (i.e. higher $m_1$), while the bottom one higher $\mu_2$. The initial outer eccentricity is set to $e_2=10^{-6}$, which remains approximately constant during the evolution 
since $\mu_2\ll 1$. This condition also implies that the evolution is well approximated by the quadrupole term of the Hamiltonian. Thus Eq. \eqref{LK} is quite accurate and the outer argument of pericenter needs not be accounted for as the quadrupole Hamiltonian is independent of it (the so-called "happy coincidence" \citep{lidov1976}). The ideal zone in the parameter space, where the triple is Hill stable and KL oscillations may occur in LISA observations, is the highlighted yellow area between the green and the blue curves. This region is larger in the case of the left panels. More specifically, in what follows we focus on the top left, where $m_1=10^5\mbox{ }\Msun$ and $m_2=10\mbox{ }\Msun$ and where the GR and GW timescales are slightly longer than in the bottom left. Further decreasing $\mu_2$ would also decrease the GR timescale, allowing KL to pump the eccentricity higher, but $\mu_2<10^{-8}$ with $m_3=10^8\mbox{ }\Msun$ would result in unphysically low compact object masses.
Most of the yellow zone is of little use, though, because high $a_1$ gives weak GW signal for sources outside of the Milky Way. For this reason, in what follows we restrict the inner semi-major axis to the range $a_1\in [1;10]\mbox{ AU}$, i.e. $a_1/R_\mathrm{g1}$ between $\sim$ $10^3$ and $10^4$.

Since the triple system under investigation takes place in a nuclear star cluster, we calculate the relevant timescales of its interactions with the surrounding objects. For the sake of simplicity, we consider uniform masses for the cluster members, $m_*=1\mbox{ }\Msun$. Assuming that the cluster is virialized, the kinetic energy of the cluster stars in the vicinity of the outer binary is $m_*\sigma^2=Gm_3m_*/a_2$. Comparing it with the total energy of the inner binary, $Gm_1m_2/(2a_1)$, 
we find that the inner binary is hard if 
\begin{equation}
    a_1 \leq a_{\rm 1, hard}=\frac{m_1m_2}{m_3m_*}a_2 
\end{equation}
or equivalently if
\begin{equation}
    \frac{a_1/R_{\rm g1}}{a_2/R_{\rm g3}} \leq \frac{m_1m_2}{(m_1+m_2)m_*}
    \approx \frac{m_2}{m_*},
\end{equation}
where the last approximate equality holds in the limit $m_2\ll m_1$. In what follows we will highlight the representative case of $m_1=10^5\mbox{ }\Msun$, $m_2=10\mbox{ }\Msun$, $m_3=10^8\mbox{ }\Msun$ and $a_2=400$ AU (see the caption of Fig. \ref{heatmap}), for which $a_{1,\rm hard}=4\mbox{ AU}$. According to Heggie's law \citep{heggie}, these binaries get even harder due to the flybys of the surrounding stars, while those which are soft ($a_1>a_\mathrm{1,hard}$) get even softer until they finally evaporate. The characteristic timescales of these processes are \citep{binneytremaine}
\begin{multline}
    T_\mathrm{hard}=\frac{\sigma}{7.6Gnm_*a_1}\frac{m_2}{m_1}
    =\frac{1}{7.6\sqrt{G}n_0}\frac{m_3^{1/2}m_2}{m_*m_1a_1a_2^{1/2-\gamma}} \approx \\ \approx 1852\mbox{ years},   
\end{multline}
\begin{multline}
    T_\mathrm{ev}=\frac{\sqrt{3}\sigma(m_1+m_2)}{32\sqrt{\pi}Gnm_*^2a_1\ln \Lambda} = \\
    =\frac{\sqrt{3}}{32\sqrt{\pi G}n_0 \ln \Lambda}\frac{\sqrt{m_3}(m_1+m_2)}{m_*^2a_1a_2^{1/2-\gamma}} \approx \\
    \approx 2.3\times10^{10}\mbox{ years},
\end{multline}
where the masses and $a_2$ are as previously and $a_1$ is set to 5 AU, the mean of its interval in our investigations, for the Coulomb logarithm we assumed $\Lambda=m_3/m_*$ \citep{alexander}, and for the number density we assumed $n=n_0 (a_2/400\mathrm{ AU})^{-\gamma}$, where $\gamma=7/4$ \citep{bahcallwolf} and $n_0=10^{9}\mbox{ pc}^{-3}$ 
so the number density is 
$\approx 10^6\mbox{ }/\mbox{pc}^3$ at 0.1 pc
\citep{neumayer}. We note that the extrapolation of the Bahcall-Wolf formula to such small distances may be inaccurate, as the number density is reduced by the central SMBH.
The distance where stars are not replenished efficiently is where the gravitational wave inspiral time into the SMBH is less than the two-body relaxation $a_2\lesssim 10^3 R_\mathrm{g3}\approx 10^3$ AU \citep{gondan}. Inside of this region the depletion of stars increases the hardening and the evaporation timescales. 

Further, \cite{deme} showed that a small population of IMBHs in the galactic nucleus perturbs the outer orbit of compact objects around the SMBH which also ultimately decreases the number of binaries in the galactic center in $\approx 10^{6}$ years. 

Binary formation through triple interactions is even less probable. Its timescale is \citep{binneytremaine}
\begin{multline}
    T_\mathrm{form}=\frac{\sigma^9}{n^2G^5m_1^5}=\frac{1}{\sqrt{G}n_0^2}\frac{m_3^{9/2}}{m_1^5a_2^{9/2-2\gamma}}\approx\\ \approx 2.43\times10^{12}\mbox{ years},
\end{multline}
which is well beyond the age of the Universe.\footnote{We note that binary formation is much more efficient in an AGN disks through the gas-capture mechanism \citep{tagawa2019} or GW capture by close encounters.}

We conclude that binary hardening, evaporation, binary disruption by IMBHs, and three-body encounters are all much longer than LISA's expected lifetime, so these effects are unlikely to take place during the observation, hence they do not directly influence our results.

\begin{figure*}
\centering
\includegraphics[scale=0.65]{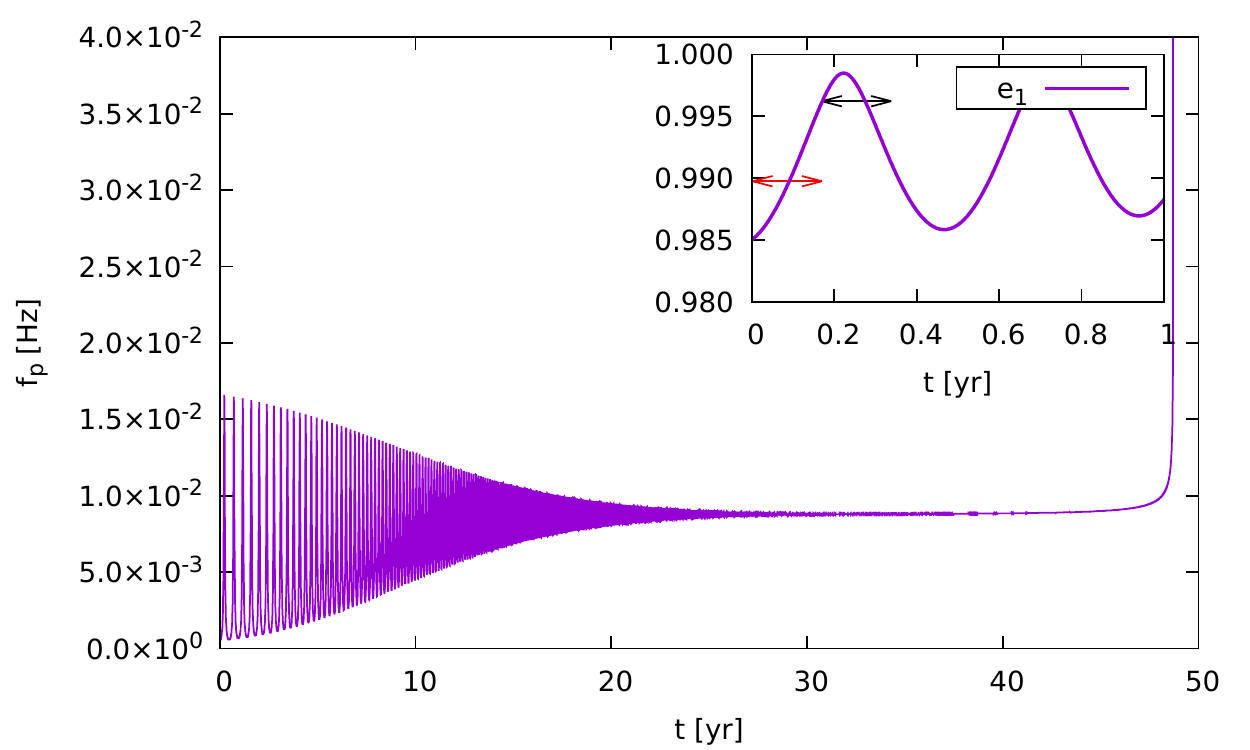}
\includegraphics[scale=0.65]{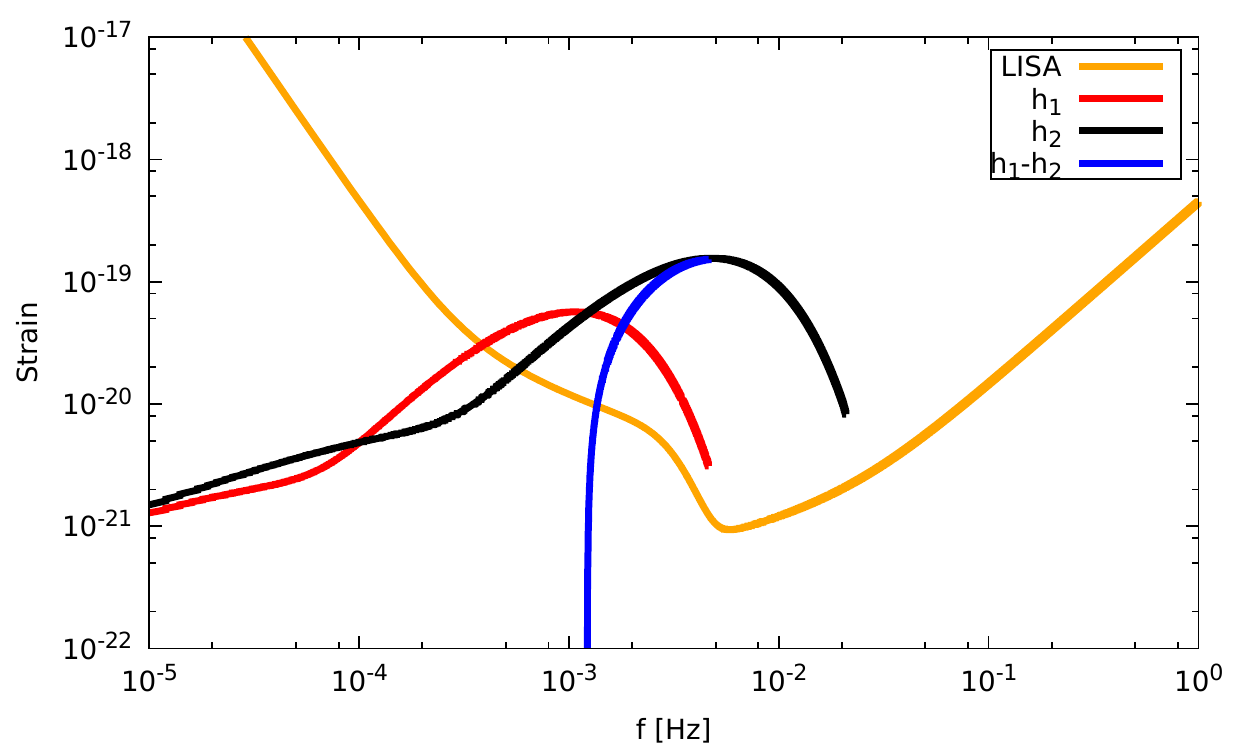}
\caption{\textit{Left}: Pericenter frequency evolution of the IMBH-SBH inner binary initiated in the parameter space from the black asterisk in Fig. \ref{heatmap}. It oscillates in the LISA frequency band due to the KL torque from the SMBH tertiary, but later becomes damped by the quick pericenter precession induced by GR. The in-set figure shows the  first two KL-peaks in the eccentricity. The horizontal black and red arrows show the interval of averaging (2 months). We use these mean values for calculating the strain (shown in black and red in the right panel). We note that the GW inspiral time is only $\sim$50 years, even though Eq. \eqref{GW1} predicts much longer values from the initial parameters ($T_\mathrm{GW,1}\approx 1.4\times 10^4$ years), because the eccentricity is increased to very high values. \textit{Right}: LISA sensitivity curve and the strain for the two 2-months intervals marked in the inset of the left panel. $h_1$ corresponds to $a_1=5.999887$ AU, $e_1=0.989745$ and $\mathrm{SNR}=5.81\,(D_{\rm L}/\Mpc)^{-1}$, while $h_2$ to $a_1=5.968728$ AU, $e_1=0.996220$ and $\mathrm{SNR}=117\,(D_{\rm L}/\Mpc)^{-1}$. The blue curve shows the difference of the strains in response to the changing orbital parameters in the inner binary, either due to the KL mechanism or GWs. 
}
\label{e1}
\end{figure*}

Fig. \ref{e1} demonstrates the time evolution of the system for a representative example shown with a star in Fig.~\ref{heatmap} at a distance of 1 Mpc. We simulate the system using the secular OSPE code. The left panel shows the pericenter frequency evolution of the inner binary. Its oscillatory behavior at the beginning is due to the KL effect, which is later quenched by GR precession. The inset of the left panel shows the first year of the inner eccentricity evolution. One way to detect the KL effect in practice is to average the GW strain over time in two-month-long intervals, indicated by horizontal arrows. We calculate the strain spectra for these averaged intervals, which are shown in the right panel of Fig. \ref{e1}. In order to detect the KL oscillations, both the GW spectral amplitude (black and red curves) and its variation (blue curve) are required to be above the LISA sensitivity curve (denoted by orange). Technically, by the difference of the strains we mean the strain of the difference of the GW signals obtained from two subsequent observational time segments.

\section{Signal-to-noise ratios}\label{snr}
In order to estimate the detectability of the signal within 
an observation segment of time duration $T_\mathrm{obs}$, we calculate the signal-to-noise ratio (SNR) following \citet{hoang2019}
\begin{equation}
    \mathrm{SNR}^2=4\int \frac{|\Tilde{h}|^2}{S_n}\mbox{ d}f,
\end{equation}
where $\Tilde{h}$ is the Fourier transform of the GW strain signal and $S_n\equiv S(f_n)$ is the LISA spectral noise amplitude. For short time segments that satisfy $T_{\rm GW1} \gg T_{\rm obs}$, and that the orbital time around the SMBH is sufficiently long, i.e. $T_{\rm orb 3}=2\pi [a_3^3 / G (m_1+m_2+m_3)]^{1/2}\gg T_{\rm obs}$ we may substitute the 
the strain for a fixed semsemimajor axis $a_1$ \citep{Peters1964}.

\begin{figure*}
\includegraphics[scale=0.5]{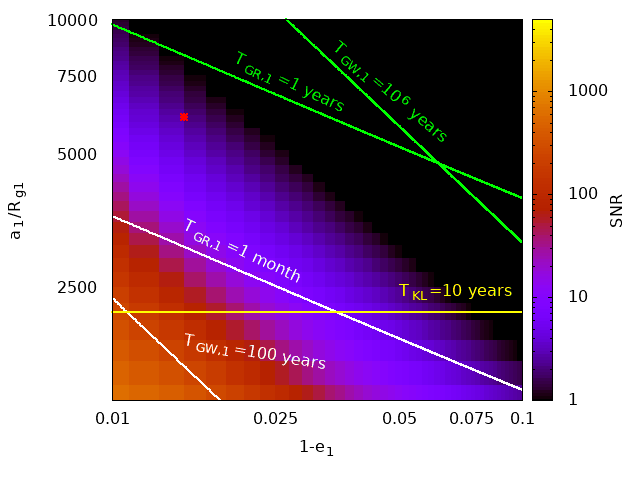}
\includegraphics[scale=0.5]{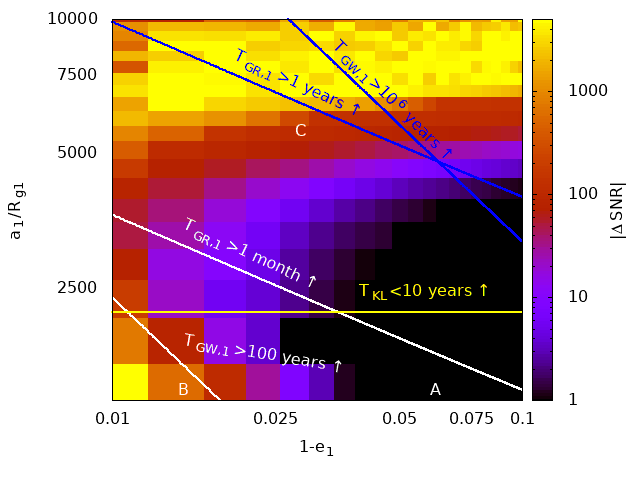}
\caption{\textit{Left}: SNR of the GW signal from the IMBH-SBH inner binary calculated from the \textit{initial} orbital parameters $e_1$ and $a_1/R_\mathrm{g1}$ for an observation time $T_\mathrm{obs}=2$ months for sources at 1 Mpc. 
The red asterisk corresponds to the same initial position as the one marked in Fig. \ref{heatmap}. We emphasize that the orbital parameters significantly change in time due to the KL oscillations, GR precession and GW radiation. By the same manner, timescales indicated by the lines reduce significantly, too.
\textit{Right}: The highest change in SNR during the evolution between two subsequent averaging time segments, maximized over both $\omega_1$ and $T_\mathrm{obs}$. In order to demonstrate the underlying dynamics we show the evolution of the orbital elements from three different parts of the parameter space, denoted by A, B and C, in Fig. \ref{evolution}.}
\label{deltasnr_ae}
\end{figure*}

The left panel of Fig. \ref{deltasnr_ae} shows the SNR for the \textit{initial} parameters of the secular evolution (calculated with $T_\mathrm{obs}=2$ months), i. e. the GW signal we would measure from the inner binary at the beginning. A red asterisk marks here the initial values used in the representative example shown in Fig. \ref{e1}. The relevant timescales for the \textit{initial} configuration are indicated with lines. However, note that these timescales change significantly during the evolution.

In order to calculate how the SNR changes during the KL evolution, we run $\sim$4000 simulations, each for 20 years and with $a_2=400$ AU, $e_2=10^{-6}$. The right panel of Fig. \ref{deltasnr_ae} indicates the maximum $\Delta$SNR, i.e. the highest change in the SNR during the evolution between two subsequent observational segments ($T_\mathrm{obs}$) for the system initiated from that particular point of the $(a_1/R_{\rm g1},1-e_1)$ parameter space. Here $\Delta$SNR is maximized over the argument of the inner pericenter $\omega_1$ in a way that it is varied in a grid from $0^\circ$ to $360^\circ$ keeping the rest of the initial elements fixed, choosing the $\omega_1$ that resulted in the highest $|\Delta\mathrm{SNR}|$. We also optimize for the $T_\mathrm{obs}$ observational time: we calculate the $\Delta$SNR for $T_\mathrm{obs}\in\{10^{-1}, 10^{0}, 10^{1}\}$ years and choose whichever gives the highest change in SNR during the evolution. We note that it makes the predicitions of the right panel somewhat pessimistic: the $\Delta$SNR values could be further increased if we chose such $T_\mathrm{obs}$ that fits better to the eccentricity oscillation timescale.  

The right panel of Fig. \ref{deltasnr_ae} shows that the high $\Delta$SNR values are found at large initial $a_1$ independently of $e_1$ and at small $a_1$ and $1-e_1$. This is not unexpected because for the former the KL time is shortest at high $a_1$ while the GR precession and inspiral time are longer there (see Eq. \eqref{LK}), therefore KL oscillations are less damped there. Interestingly, in this region the binary would not be detected without the KL oscillations, which push the binary to high eccentricities. For small $a_1$ and high $e_1$, the orbital parameters change rapidly due to the GW inspiral independently of the KL effect, which explains the lower left peak of $\Delta$SNR in Fig. \ref{deltasnr_ae}.

\begin{figure*}
\includegraphics[scale=0.7]{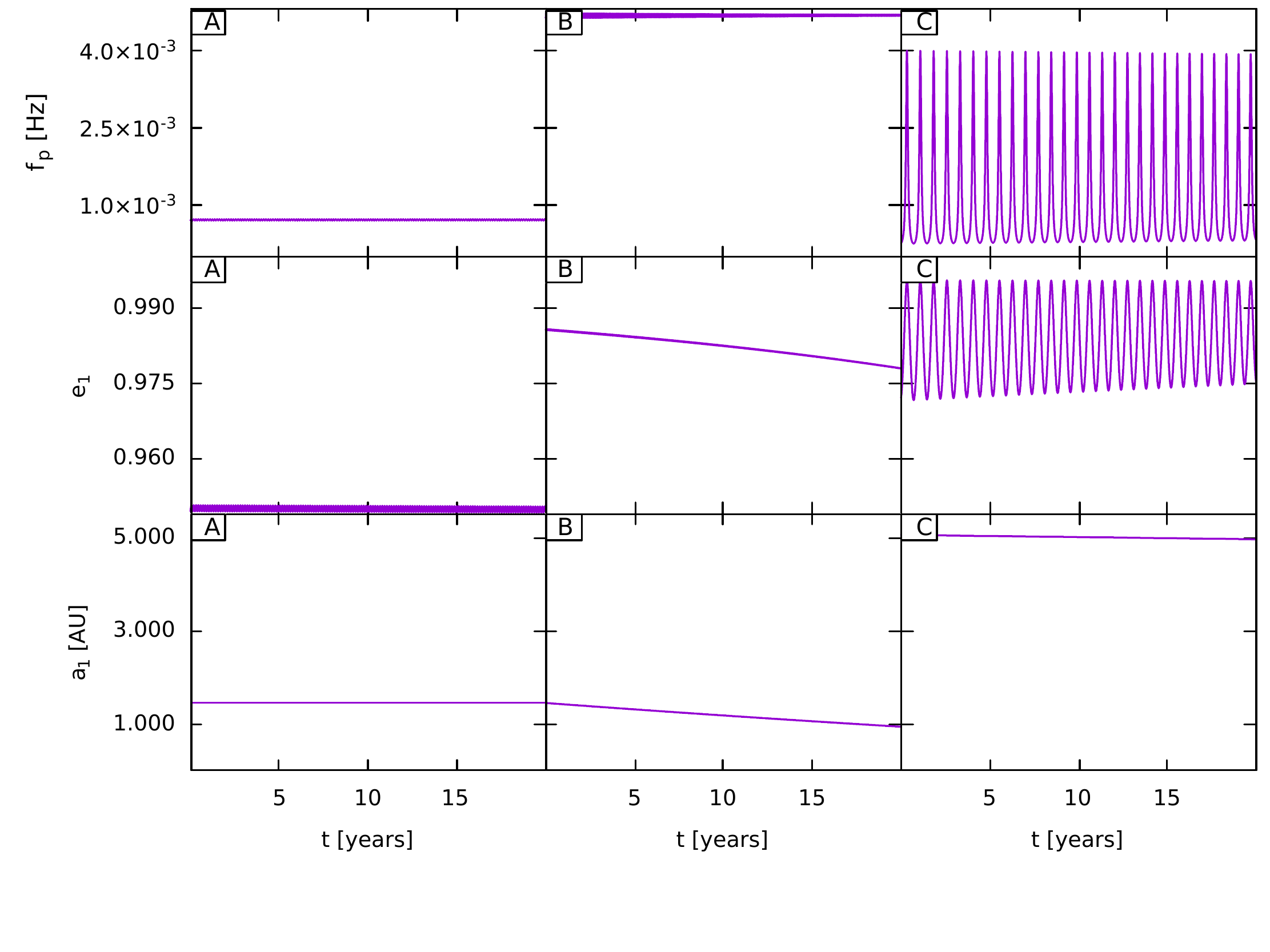}
\caption{The orbital element evolution of the three systems sampled from the right panel of Fig. \ref{deltasnr_ae}. Evolution A and B are both highly damped by GR. The high $\Delta$SNR value at B is the result of the strong GW decay.}
\label{evolution}
\end{figure*}

\begin{figure*}
\includegraphics[scale=0.9]{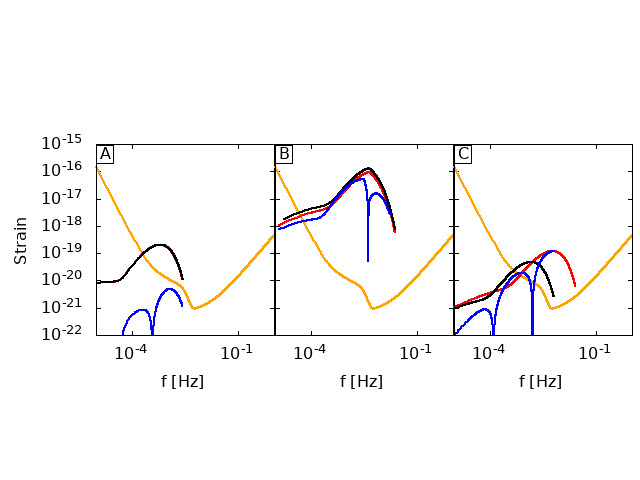}
\caption{The GW spectra of the systems sampled from the right panel of Fig. \ref{deltasnr_ae}. The black and red curves correspond to the signals which give the highest $\Delta$SNR during the evolution. The blue curve is the difference of them. In case A the black and red curves are so close to each other that they are not separable by eye.}
\label{evolution_strain}
\end{figure*}

To better understand this behavior, we select three representative points (denoted by A, B and C in Fig.~\ref{deltasnr_ae}) and plot the time evolution of their orbital elements in Fig. \ref{evolution} and the GW spectral amplitude in Fig.~\ref{evolution_strain}. The first row of panels shows the pericenter frequency calculated as
\begin{equation}\label{eq:fp}
    f_\mathrm{p}=\frac{1}{2\pi}\sqrt{\frac{G(m_1+m_2)}{a_1^3}\frac{(1+e_1)}{(1-e_1)^3}}.
\end{equation}
The figures show that Case C exhibits multiple prominent KL oscillation cycles, which leads to a high $\Delta$SNR. However, in Case B, KL oscillations are quenched by the rapid GR precession. We note that even in such a quenched case there are some small oscillations ($\Delta e \sim 10^{-4}$, $T_\mathrm{GR}\sim 10^{-2}$ years \citep[e.g.,][]{naoz2013}), but they do not produce significant $\Delta$SNR because of the small change in the eccentricity. The high $\Delta$SNR is obtained with $T_\mathrm{obs}=10\yr$: the reason for this is that most of the variation of the orbital parameters is caused by the GW inspiral (not by the SMBH), so we need to have a $T_\mathrm{obs}$ that is comparable to the inspiral time, $T_{\rm GW,1}\approx 57\yr$. The high $\Delta$SNR in thus mostly independent of the KL effect. In case A, the orbital parameters are almost constant as the system is neither inspiraling nor does it exhibit KL oscillations. 

\section{Discussion and conclusion}\label{discussion}

We have shown that the dynamical imprint of SMBHs may be significantly detected with LISA from 1 Mpc for compact objects orbiting IMBHs in galactic nuclei. Fig. \ref{deltasnr_ae} showed the initial orbital parameters where this identification is possible. We found that the imprint of KL oscillations are most prominent for IMBH sources orbited by a stellar mass compact object which orbit around a SMBH. A binary of two stellar mass compact objects also exhibit similar oscillations in the vicinity of a SMBH, but in this case either the GW strain amplitude is much smaller or the GR precession rate is higher which decreases the KL oscillation amplitude. Further, KL oscillations are also less prominent in hierarchical SMBH triples since in this case the KL timescale is typically much longer than the observation time.

To demonstrate the detectability of KL oscillations in a robust way, we calculated the variations of the SNR during the observation in fixed $T_{\rm obs}$ duration segments of the total observation period, and marginalized over the value of $T_{\rm obs}$. This analysis showed that the variations due to the KL effect can be highly significant and detectable with LISA to at least 1 Mpc. 

While we have highlighted cases where the full KL oscillations may be detected with LISA with very high significance, the true parameter space where the KL effect may be detected is certainly much larger. Since the number of GW cycles is of order $N_{\rm GW}\sim T_{\rm obs} f_{\rm orb} \sim T_{\rm obs}f_{\rm p} (1-e_1)^{3/2}$ (Eq.~\ref{eq:fp}), a very small variation of eccentricity of order 
\begin{equation}
4\times 10^{-4} 
\left(\frac{T_{\rm obs}}{4\rm yr}\right)^{-2/3} \left(\frac{f_{\rm p}}{1 \mathrm{mHz}}\right)^{-2/3}   
\end{equation}
may cause order unity change in the number of detected cycles during a 4 year observation. Thus, the KL effect may be significant even if only a $10^{-3}$ fraction of a full KL cycle is observed. Furthermore, KL oscillations may push the binary to so high eccentricities that the binary merges during the observation. For merging binaries, the number of GW cycles is proportional to the inverse GW timescale, which for asymptotically high $e_1$ close to unity is proportional to $(1-e_1^2)^{-7/2}$ (Eq.~\ref{GW1}), implying an even higher sensitivity to eccentricity. Thus, the KL effect of inspiraling GW sources may be highly significant even in cases where the observation time and/or the GW inspiral time is much shorter than the KL timescale. 

While we leave the detailed GW data analysis exploration of KL imprints to a future study, these arguments suggest that the detection prospects of the KL effect may be possible even beyond the case of IMBH-stellar mass compact object triples around SMBHs.

\section*{Acknowledgements}
This project has received funding from the European Research Council (ERC) under the European Union's Horizon 2020 research and innovation programme under grant agreement No 638435 (GalNUC) and by the Hungarian National Research, Development, and Innovation Office grant NKFIH KH-125675. B.M.H. and S.N. acknowledge the partial support of NASA grants Nos. 80NSSC19K0321 and 80NSSC20K0505. S.N. also thanks Howard and Astrid Preston for their generous support.

\bibliography{cit}{}

\begin{thebibliography}{}
\expandafter\ifx\csname natexlab\endcsname\relax\def\natexlab#1{#1}\fi
\providecommand{\url}[1]{\href{#1}{#1}}
\providecommand{\dodoi}[1]{doi:~\href{http://doi.org/#1}{\nolinkurl{#1}}}
\providecommand{\doeprint}[1]{\href{http://ascl.net/#1}{\nolinkurl{http://ascl.net/#1}}}
\providecommand{\doarXiv}[1]{\href{https://arxiv.org/abs/#1}{\nolinkurl{https://arxiv.org/abs/#1}}}

\bibitem[{{Abbott} {et~al.}(2016){Abbott}, {Abbott}, {Abbott}, {Abernathy},
  {Acernese}, {Ackley}, {Adams}, {Adams}, {Addesso}, {Adhikari}, {Adya},
  {Affeldt}, {Agathos}, {Agatsuma}, {Aggarwal}, {Aguiar}, {Aiello}, {Ain},
  {Ajith}, {Allen}, {Allocca}, {Altin}, {Anderson}, {Anderson}, {Arai},
  {Araya}, {Arceneaux}, {Areeda}, {Arnaud}, {Arun}, {Ascenzi}, {Ashton}, {Ast},
  {Aston}, {Astone}, {Aufmuth}, {Aulbert}, {Babak}, {Bacon}, {Bader}, {Baker},
  {Baldaccini}, {Ballardin}, {Ballmer}, {Barayoga}, {Barclay}, {Barish},
  {Barker}, {Barone}, {Barr}, {Barsotti}, {Barsuglia}, {Barta}, {Bartlett},
  {Bartos}, {Bassiri}, {Basti}, {Batch}, {Baune}, {Bavigadda}, {Bazzan},
  {Bejger}, {Bell}, {Berger}, {Bergmann}, {Berry}, {Bersanetti}, {Bertolini},
  {Betzwieser}, {Bhagwat}, {Bhandare}, {Bilenko}, {Billingsley}, {Birch},
  {Birney}, {Birnholtz}, {Biscans}, {Bisht}, {Bitossi}, {Biwer}, {Bizouard},
  {Blackburn}, {Blair}, {Blair}, {Blair}, {Bloemen}, {Bock}, {Boer}, {Bogaert},
  {Bogan}, {Bohe}, {Bond}, {Bondu}, {Bonnand}, {Boom}, {Bork}, {Boschi},
  {Bose}, {Bouffanais}, {Bozzi}, {Bradaschia}, {Brady}, {Braginsky},
  {Branchesi}, {Brau}, {Briant}, {Brillet}, {Brinkmann}, {Brisson}, {Brockill},
  {Broida}, {Brooks}, {Brown}, {Brown}, {Brown}, {Brunett}, {Buchanan},
  {Buikema}, {Bulik}, {Bulten}, {Buonanno}, {Buskulic}, {Buy}, {Byer},
  {Cabero}, {Cadonati}, {Cagnoli}, {Cahillane}, {Calder{\'o}n Bustillo},
  {Callister}, {Calloni}, {Camp}, {Cannon}, {Cao}, {Capano}, {Capocasa},
  {Carbognani}, {Caride}, {Casanueva Diaz}, {Casentini}, {Caudill},
  {Cavagli{\`a}}, {Cavalier}, {Cavalieri}, {Cella}, {Cepeda}, {Cerboni
  Baiardi}, {Cerretani}, {Cesarini}, {Chamberlin}, {Chan}, {Chao}, {Charlton},
  {Chassande-Mottin}, {Cheeseboro}, {Chen}, {Chen}, {Cheng}, {Chincarini},
  {Chiummo}, {Cho}, {Cho}, {Chow}, {Christensen}, {Chu}, {Chua}, {Chung},
  {Ciani}, {Clara}, {Clark}, {Cleva}, {Coccia}, {Cohadon}, {Colla}, {Collette},
  {Cominsky}, {Constancio}, {Conte}, {Conti}, {Cook}, {Corbitt}, {Cornish},
  {Corsi}, {Cortese}, {Costa}, {Coughlin}, {Coughlin}, {Coulon}, {Countryman},
  {Couvares}, {Cowan}, {Coward}, {Cowart}, {Coyne}, {Coyne}, {Craig},
  {Creighton}, {Cripe}, {Crowder}, {Cumming}, {Cunningham}, {Cuoco}, {Dal
  Canton}, {Danilishin}, {D'Antonio}, {Danzmann}, {Darman}, {Dasgupta}, {Da
  Silva Costa}, {Dattilo}, {Dave}, {Davier}, {Davies}, {Daw}, {Day}, {De},
  {DeBra}, {Debreczeni}, {Degallaix}, {De Laurentis}, {Del{\'e}glise}, {Del
  Pozzo}, {Denker}, {Dent}, {Dergachev}, {De Rosa}, {DeRosa}, {DeSalvo},
  {Devine}, {Dhurand har}, {D{\'\i}az}, {Di Fiore}, {Di Giovanni}, {Di
  Girolamo}, {Di Lieto}, {Di Pace}, {Di Palma}, {Di Virgilio}, {Dolique},
  {Donovan}, {Dooley}, {Doravari}, {Douglas}, {Downes}, {Drago}, {Drever},
  {Driggers}, {Ducrot}, {Dwyer}, {Edo}, {Edwards}, {Effler}, {Eggenstein},
  {Ehrens}, {Eichholz}, {Eikenberry}, {Engels}, {Essick}, {Etzel}, {Evans},
  {Evans}, {Everett}, {Factourovich}, {Fafone}, {Fair}, {Fairhurst}, {Fan},
  {Fang}, {Farinon}, {Farr}, {Farr}, {Favata}, {Fays}, {Fehrmann}, {Fejer},
  {Fenyvesi}, {Ferrante}, {Ferreira}, {Ferrini}, {Fidecaro}, {Fiori},
  {Fiorucci}, {Fisher}, {Flaminio}, {Fletcher}, {Fong}, {Fournier}, {Frasca},
  {Frasconi}, {Frei}, {Freise}, {Frey}, {Frey}, {Fritschel}, {Frolov}, {Fulda},
  {Fyffe}, {Gabbard}, {Gaebel}, {Gair}, {Gammaitoni}, {Gaonkar}, {Garufi},
  {Gaur}, {Gehrels}, {Gemme}, {Geng}, {Genin}, {Gennai}, {George}, {Gergely},
  {Germain}, {Ghosh}, {Ghosh}, {Ghosh}, {Giaime}, {Giardina}, {Giazotto},
  {Gill}, {Glaefke}, {Goetz}, {Goetz}, {Gondan}, {Gonz{\'a}lez}, {Gonzalez
  Castro}, {Gopakumar}, {Gordon}, {Gorodetsky}, {Gossan}, {Gosselin}, {Gouaty},
  {Grado}, {Graef}, {Graff}, {Granata}, {Grant}, {Gras}, {Gray}, {Greco},
  {Green}, {Groot}, {Grote}, {Grunewald}, {Guidi}, {Guo}, {Gupta}, {Gupta},
  {Gushwa}, {Gustafson}, {Gustafson}, {Hacker}, {Hall}, {Hall}, {Hamilton},
  {Hammond}, {Haney}, {Hanke}, {Hanks}, {Hanna}, {Hannam}, {Hanson},
  {Hardwick}, {Harms}, {Harry}, {Harry}, {Hart}, {Hartman}, {Haster},
  {Haughian}, {Healy}, {Heidmann}, {Heintze}, {Heitmann}, {Hello}, {Hemming},
  {Hendry}, {Heng}, {Hennig}, {Henry}, {Heptonstall}, {Heurs}, {Hild}, {Hoak},
  {Hofman}, {Holt}, {Holz}, {Hopkins}, {Hough}, {Houston}, {Howell}, {Hu},
  {Huang}, {Huerta}, {Huet}, {Hughey}, {Husa}, {Huttner}, {Huynh-Dinh},
  {Indik}, {Ingram}, {Inta}, {Isa}, {Isac}, {Isi}, {Isogai}, {Iyer}, {Izumi},
  {Jacqmin}, {Jang}, {Jani}, {Jaranowski}, {Jawahar}, {Jian},
  {Jim{\'e}nez-Forteza}, {Johnson}, {Johnson-McDaniel}, {Jones}, {Jones},
  {Jonker}, {Ju}, {K}, {Kalaghatgi}, {Kalogera}, {Kandhasamy}, {Kang},
  {Kanner}, {Kapadia}, {Karki}, {Karvinen}, {Kasprzack}, {Katsavounidis},
  {Katzman}, {Kaufer}, {Kaur}, {Kawabe}, {K{\'e}f{\'e}lian}, {Kehl}, {Keitel},
  {Kelley}, {Kells}, {Kennedy}, {Key}, {Khalili}, {Khan}, {Khan}, {Khan},
  {Khazanov}, {Kijbunchoo}, {Kim}, {Kim}, {Kim}, {Kim}, {Kim}, {Kim}, {Kim},
  {Kimbrell}, {King}, {King}, {Kissel}, {Klein}, {Kleybolte}, {Klimenko},
  {Koehlenbeck}, {Koley}, {Kondrashov}, {Kontos}, {Korobko}, {Korth},
  {Kowalska}, {Kozak}, {Kringel}, {Krishnan}, {Kr{\'o}lak}, {Krueger}, {Kuehn},
  {Kumar}, {Kumar}, {Kuo}, {Kutynia}, {Lackey}, {Land ry}, {Lange}, {Lantz},
  {Lasky}, {Laxen}, {Lazzarini}, {Lazzaro}, {Leaci}, {Leavey}, {Lebigot},
  {Lee}, {Lee}, {Lee}, {Lee}, {Lenon}, {Leonardi}, {Leong}, {Leroy},
  {Letendre}, {Levin}, {Lewis}, {Li}, {Libson}, {Littenberg}, {Lockerbie},
  {Lombardi}, {London}, {Lord}, {Lorenzini}, {Loriette}, {Lormand}, {Losurdo},
  {Lough}, {Lousto}, {L{\"u}ck}, {Lundgren}, {Lynch}, {Ma}, {Machenschalk},
  {MacInnis}, {Macleod}, {Maga{\~n}a-Sandoval}, {Maga{\~n}a Zertuche}, {Magee},
  {Majorana}, {Maksimovic}, {Malvezzi}, {Man}, {Mandel}, {Mandic}, {Mangano},
  {Mansell}, {Manske}, {Mantovani}, {Marchesoni}, {Marion}, {M{\'a}rka},
  {M{\'a}rka}, {Markosyan}, {Maros}, {Martelli}, {Martellini}, {Martin},
  {Martynov}, {Marx}, {Mason}, {Masserot}, {Massinger}, {Masso-Reid},
  {Mastrogiovanni}, {Matichard}, {Matone}, {Mavalvala}, {Mazumder}, {McCarthy},
  {McClelland}, {McCormick}, {McGuire}, {McIntyre}, {McIver}, {McManus},
  {McRae}, {McWilliams}, {Meacher}, {Meadors}, {Meidam}, {Melatos}, {Mendell},
  {Mercer}, {Merilh}, {Merzougui}, {Meshkov}, {Messenger}, {Messick},
  {Metzdorff}, {Meyers}, {Mezzani}, {Miao}, {Michel}, {Middleton}, {Mikhailov},
  {Milano}, {Miller}, {Miller}, {Miller}, {Miller}, {Millhouse}, {Minenkov},
  {Ming}, {Mirshekari}, {Mishra}, {Mitra}, {Mitrofanov}, {Mitselmakher},
  {Mittleman}, {Moggi}, {Mohan}, {Mohapatra}, {Montani}, {Moore}, {Moore},
  {Moraru}, {Moreno}, {Morriss}, {Mossavi}, {Mours}, {Mow-Lowry}, {Mueller},
  {Muir}, {Mukherjee}, {Mukherjee}, {Mukherjee}, {Mukund}, {Mullavey}, {Munch},
  {Murphy}, {Murray}, {Mytidis}, {Nardecchia}, {Naticchioni}, {Nayak},
  {Nedkova}, {Nelemans}, {Nelson}, {Neri}, {Neunzert}, {Newton}, {Nguyen},
  {Nielsen}, {Nissanke}, {Nitz}, {Nocera}, {Nolting}, {Normandin}, {Nuttall},
  {Oberling}, {Ochsner}, {O'Dell}, {Oelker}, {Ogin}, {Oh}, {Oh}, {Ohme},
  {Oliver}, {Oppermann}, {Oram}, {O'Reilly}, {O'Shaughnessy}, {Ottaway},
  {Overmier}, {Owen}, {Pai}, {Pai}, {Palamos}, {Palashov}, {Palomba},
  {Pal-Singh}, {Pan}, {Pan}, {Pankow}, {Pannarale}, {Pant}, {Paoletti},
  {Paoli}, {Papa}, {Paris}, {Parker}, {Pascucci}, {Pasqualetti}, {Passaquieti},
  {Passuello}, {Patricelli}, {Patrick}, {Pearlstone}, {Pedraza}, {Pedurand},
  {Pekowsky}, {Pele}, {Penn}, {Perreca}, {Perri}, {Pfeiffer}, {Phelps},
  {Piccinni}, {Pichot}, {Piergiovanni}, {Pierro}, {Pillant}, {Pinard}, {Pinto},
  {Pitkin}, {Poe}, {Poggiani}, {Popolizio}, {Porter}, {Post}, {Powell},
  {Prasad}, {Predoi}, {Prestegard}, {Price}, {Prijatelj}, {Principe},
  {Privitera}, {Prix}, {Prodi}, {Prokhorov}, {Puncken}, {Punturo}, {Puppo},
  {P{\"u}rrer}, {Qi}, {Qin}, {Qiu}, {Quetschke}, {Quintero}, {Quitzow-James},
  {Raab}, {Rabeling}, {Radkins}, {Raffai}, {Raja}, {Rajan}, {Rakhmanov},
  {Rapagnani}, {Raymond}, {Razzano}, {Re}, {Read}, {Reed}, {Regimbau}, {Rei},
  {Reid}, {Reitze}, {Rew}, {Reyes}, {Ricci}, {Riles}, {Rizzo}, {Robertson},
  {Robie}, {Robinet}, {Rocchi}, {Rolland}, {Rollins}, {Roma}, {Romano},
  {Romano}, {Romanov}, {Romie}, {Rosi{\'n}ska}, {Rowan}, {R{\"u}diger},
  {Ruggi}, {Ryan}, {Sachdev}, {Sadecki}, {Sadeghian}, {Sakellariadou},
  {Salconi}, {Saleem}, {Salemi}, {Samajdar}, {Sammut}, {Sanchez}, {Sandberg},
  {Sandeen}, {Sand ers}, {Sassolas}, {Sathyaprakash}, {Saulson}, {Sauter},
  {Savage}, {Sawadsky}, {Schale}, {Schilling}, {Schmidt}, {Schmidt},
  {Schnabel}, {Schofield}, {Sch{\"o}nbeck}, {Schreiber}, {Schuette}, {Schutz},
  {Scott}, {Scott}, {Sellers}, {Sengupta}, {Sentenac}, {Sequino}, {Sergeev},
  {Setyawati}, {Shaddock}, {Shaffer}, {Shahriar}, {Shaltev}, {Shapiro},
  {Shawhan}, {Sheperd}, {Shoemaker}, {Shoemaker}, {Siellez}, {Siemens},
  {Sieniawska}, {Sigg}, {Silva}, {Singer}, {Singer}, {Singh}, {Singh},
  {Singhal}, {Sintes}, {Slagmolen}, {Smith}, {Smith}, {Smith}, {Son}, {Sorazu},
  {Sorrentino}, {Souradeep}, {Srivastava}, {Staley}, {Steinke}, {Steinlechner},
  {Steinlechner}, {Steinmeyer}, {Stephens}, {Stevenson}, {Stone}, {Strain},
  {Straniero}, {Stratta}, {Strauss}, {Strigin}, {Sturani}, {Stuver},
  {Summerscales}, {Sun}, {Sunil}, {Sutton}, {Swinkels}, {Szczepa{\'n}czyk},
  {Tacca}, {Talukder}, {Tanner}, {T{\'a}pai}, {Tarabrin}, {Taracchini},
  {Taylor}, {Theeg}, {Thirugnanasamband am}, {Thomas}, {Thomas}, {Thomas},
  {Thorne}, {Thrane}, {Tiwari}, {Tiwari}, {Tokmakov}, {Toland}, {Tomlinson},
  {Tonelli}, {Tornasi}, {Torres}, {Torrie}, {T{\"o}yr{\"a}}, {Travasso},
  {Traylor}, {Trifir{\`o}}, {Tringali}, {Trozzo}, {Tse}, {Turconi},
  {Tuyenbayev}, {Ugolini}, {Unnikrishnan}, {Urban}, {Usman}, {Vahlbruch},
  {Vajente}, {Valdes}, {Vallisneri}, {van Bakel}, {van Beuzekom}, {van den
  Brand}, {Van Den Broeck}, {Vand er-Hyde}, {van der Schaaf}, {van Heijningen},
  {van Veggel}, {Vardaro}, {Vass}, {Vas{\'u}th}, {Vaulin}, {Vecchio},
  {Vedovato}, {Veitch}, {Veitch}, {Venkateswara}, {Verkindt}, {Vetrano},
  {Vicer{\'e}}, {Vinciguerra}, {Vine}, {Vinet}, {Vitale}, {Vo}, {Vocca},
  {Vorvick}, {Voss}, {Vousden}, {Vyatchanin}, {Wade}, {Wade}, {Wade}, {Walker},
  {Wallace}, {Walsh}, {Wang}, {Wang}, {Wang}, {Wang}, {Wang}, {Ward}, {Warner},
  {Was}, {Weaver}, {Wei}, {Weinert}, {Weinstein}, {Weiss}, {Wen}, {We{\ss}els},
  {Westphal}, {Wette}, {Whelan}, {Whitcomb}, {Whiting}, {Williams},
  {Williamson}, {Willis}, {Willke}, {Wimmer}, {Winkler}, {Wipf}, {Wittel},
  {Woan}, {Woehler}, {Worden}, {Wright}, {Wu}, {Wu}, {Yablon}, {Yam},
  {Yamamoto}, {Yancey}, {Yu}, {Yvert}, {Zadro{\.Z}ny}, {Zangrando}, {Zanolin},
  {Zendri}, {Zevin}, {Zhang}, {Zhang}, {Zhang}, {Zhao}, {Zhou}, {Zhou}, {Zhu},
  {Zucker}, {Zuraw}, {Zweizig}, {LIGO Scientific Collaboration}, \& {Virgo
  Collaboration}}]{abbott2016}
{Abbott}, B.~P., {Abbott}, R., {Abbott}, T.~D., {et~al.} 2016, Physical Review
  X, 6, 041015, \dodoi{10.1103/PhysRevX.6.041015}

\bibitem[{{Alexander}(2017)}]{alexander}
{Alexander}, T. 2017, \araa, 55, 17,
  \dodoi{10.1146/annurev-astro-091916-055306}

\bibitem[{{Amaro-Seoane} {et~al.}(2017){Amaro-Seoane}, {Audley}, {Babak},
  {Baker}, {Barausse}, {Bender}, {Berti}, {Binetruy}, {Born}, {Bortoluzzi},
  {Camp}, {Caprini}, {Cardoso}, {Colpi}, {Conklin}, {Cornish}, {Cutler},
  {Danzmann}, {Dolesi}, {Ferraioli}, {Ferroni}, {Fitzsimons}, {Gair}, {Gesa
  Bote}, {Giardini}, {Gibert}, {Grimani}, {Halloin}, {Heinzel}, {Hertog},
  {Hewitson}, {Holley-Bockelmann}, {Hollington}, {Hueller}, {Inchauspe},
  {Jetzer}, {Karnesis}, {Killow}, {Klein}, {Klipstein}, {Korsakova}, {Larson},
  {Livas}, {Lloro}, {Man}, {Mance}, {Martino}, {Mateos}, {McKenzie},
  {McWilliams}, {Miller}, {Mueller}, {Nardini}, {Nelemans}, {Nofrarias},
  {Petiteau}, {Pivato}, {Plagnol}, {Porter}, {Reiche}, {Robertson},
  {Robertson}, {Rossi}, {Russano}, {Schutz}, {Sesana}, {Shoemaker}, {Slutsky},
  {Sopuerta}, {Sumner}, {Tamanini}, {Thorpe}, {Troebs}, {Vallisneri},
  {Vecchio}, {Vetrugno}, {Vitale}, {Volonteri}, {Wanner}, {Ward}, {Wass},
  {Weber}, {Ziemer}, \& {Zweifel}}]{amaro-seoane}
{Amaro-Seoane}, P., {Audley}, H., {Babak}, S., {et~al.} 2017, arXiv e-prints,
  arXiv:1702.00786.
\newblock \doarXiv{1702.00786}

\bibitem[{{Antonini} {et~al.}(2014){Antonini}, {Murray}, \&
  {Mikkola}}]{antonini2014}
{Antonini}, F., {Murray}, N., \& {Mikkola}, S. 2014, \apj, 781, 45,
  \dodoi{10.1088/0004-637X/781/1/45}

\bibitem[{{Bahcall} \& {Wolf}(1976)}]{bahcallwolf}
{Bahcall}, J.~N., \& {Wolf}, R.~A. 1976, \apj, 209, 214, \dodoi{10.1086/154711}

\bibitem[{{Begelman} {et~al.}(1980){Begelman}, {Blandford}, \&
  {Rees}}]{begelman}
{Begelman}, M.~C., {Blandford}, R.~D., \& {Rees}, M.~J. 1980, Nature, 287, 307,
  \dodoi{10.1038/287307a0}

\bibitem[{{Binney} \& {Tremaine}(2008)}]{binneytremaine}
{Binney}, J., \& {Tremaine}, S. 2008, {Galactic Dynamics: Second Edition}

\bibitem[{{Blecha} \& {Loeb}(2008)}]{Blecha+08}
{Blecha}, L., \& {Loeb}, A. 2008, MNRAS, 390, 1311,
  \dodoi{10.1111/j.1365-2966.2008.13790.x}

\bibitem[{{Chen} \& {Liu}(2013)}]{Chen+13}
{Chen}, X., \& {Liu}, F.~K. 2013, \apj, 762, 95,
  \dodoi{10.1088/0004-637X/762/2/95}

\bibitem[{{Chen} {et~al.}(2009){Chen}, {Madau}, {Sesana}, \& {Liu}}]{Chen+09}
{Chen}, X., {Madau}, P., {Sesana}, A., \& {Liu}, F.~K. 2009, \apjl, 697, L149,
  \dodoi{10.1088/0004-637X/697/2/L149}

\bibitem[{{Chen} {et~al.}(2011){Chen}, {Sesana}, {Madau}, \& {Liu}}]{Chen+11}
{Chen}, X., {Sesana}, A., {Madau}, P., \& {Liu}, F.~K. 2011, \apj, 729, 13,
  \dodoi{10.1088/0004-637X/729/1/13}

\bibitem[{{Deme} {et~al.}(2020){Deme}, {Meiron}, \& {Kocsis}}]{deme}
{Deme}, B., {Meiron}, Y., \& {Kocsis}, B. 2020, \apj, 892, 130,
  \dodoi{10.3847/1538-4357/ab7921}

\bibitem[{{Di Matteo} {et~al.}(2005){Di Matteo}, {Springel}, \&
  {Hernquist}}]{DiMatteo+05}
{Di Matteo}, T., {Springel}, V., \& {Hernquist}, L. 2005, \nat, 433, 604,
  \dodoi{10.1038/nature03335}

\bibitem[{{Emami} \& {Loeb}(2019)}]{emami2019}
{Emami}, R., \& {Loeb}, A. 2019, arXiv e-prints, arXiv:1910.04828.
\newblock \doarXiv{1910.04828}

\bibitem[{{Fang} \& {Huang}(2020)}]{yun2020}
{Fang}, Y., \& {Huang}, Q.-G. 2020, arXiv e-prints, arXiv:2004.09390.
\newblock \doarXiv{2004.09390}

\bibitem[{{Fragione} \& {Gualandris}(2019)}]{Fragione+19}
{Fragione}, G., \& {Gualandris}, A. 2019, \mnras, 489, 4543,
  \dodoi{10.1093/mnras/stz2451}

\bibitem[{{Fragione} {et~al.}(2020){Fragione}, {Loeb}, {Kremer}, \&
  {Rasio}}]{Fragione+20}
{Fragione}, G., {Loeb}, A., {Kremer}, K., \& {Rasio}, F.~A. 2020, arXiv
  e-prints, arXiv:2002.02975.
\newblock \doarXiv{2002.02975}

\bibitem[{{Genzel} {et~al.}(2010){Genzel}, {Eisenhauer}, \&
  {Gillessen}}]{Genzel+10}
{Genzel}, R., {Eisenhauer}, F., \& {Gillessen}, S. 2010, Reviews of Modern
  Physics, 82, 3121, \dodoi{10.1103/RevModPhys.82.3121}

\bibitem[{{Ghez} {et~al.}(2008){Ghez}, {Salim}, {Weinberg}, {Lu}, {Do}, {Dunn},
  {Matthews}, {Morris}, {Yelda}, {Becklin}, {Kremenek}, {Milosavljevic}, \&
  {Naiman}}]{Ghez+08}
{Ghez}, A.~M., {Salim}, S., {Weinberg}, N.~N., {et~al.} 2008, \apj, 689, 1044,
  \dodoi{10.1086/592738}

\bibitem[{{Gond{\'a}n} {et~al.}(2018){Gond{\'a}n}, {Kocsis}, {Raffai}, \&
  {Frei}}]{gondan}
{Gond{\'a}n}, L., {Kocsis}, B., {Raffai}, P., \& {Frei}, Z. 2018, \apj, 860, 5,
  \dodoi{10.3847/1538-4357/aabfee}

\bibitem[{{Goodman} \& {Tan}(2004)}]{Goodman_Tan2004}
{Goodman}, J., \& {Tan}, J.~C. 2004, \apj, 608, 108, \dodoi{10.1086/386360}

\bibitem[{{Greene} {et~al.}(2019){Greene}, {Strader}, \& {Ho}}]{greene}
{Greene}, J.~E., {Strader}, J., \& {Ho}, L.~C. 2019, arXiv e-prints,
  arXiv:1911.09678.
\newblock \doarXiv{1911.09678}

\bibitem[{{Gualandris} {et~al.}(2010){Gualandris}, {Gillessen}, \&
  {Merritt}}]{Gualandris+2010}
{Gualandris}, A., {Gillessen}, S., \& {Merritt}, D. 2010, \mnras, 409, 1146,
  \dodoi{10.1111/j.1365-2966.2010.17373.x}

\bibitem[{{Gualandris} \& {Merritt}(2009)}]{Gualandris_Merritt2009}
{Gualandris}, A., \& {Merritt}, D. 2009, \apj, 705, 361,
  \dodoi{10.1088/0004-637X/705/1/361}

\bibitem[{{Gualandris} \& {Merritt}(2012)}]{Gualandris_Merritt2012}
---. 2012, \apj, 744, 74, \dodoi{10.1088/0004-637X/744/1/74}

\bibitem[{{Gupta} {et~al.}(2019){Gupta}, {Suzuki}, {Okawa}, \&
  {Maeda}}]{gupta2019}
{Gupta}, P., {Suzuki}, H., {Okawa}, H., \& {Maeda}, K.-i. 2019, arXiv e-prints,
  arXiv:1911.11318.
\newblock \doarXiv{1911.11318}

\bibitem[{{Hamers} {et~al.}(2018){Hamers}, {Bar-Or}, {Petrovich}, \&
  {Antonini}}]{hamers2018}
{Hamers}, A.~S., {Bar-Or}, B., {Petrovich}, C., \& {Antonini}, F. 2018, \apj,
  865, 2, \dodoi{10.3847/1538-4357/aadae2}

\bibitem[{{Heggie}(1975)}]{heggie}
{Heggie}, D.~C. 1975, \mnras, 173, 729, \dodoi{10.1093/mnras/173.3.729}

\bibitem[{{Hoang} {et~al.}(2019){Hoang}, {Naoz}, {Kocsis}, {Farr}, \&
  {McIver}}]{hoang2019}
{Hoang}, B.-M., {Naoz}, S., {Kocsis}, B., {Farr}, W.~M., \& {McIver}, J. 2019,
  \apjl, 875, L31, \dodoi{10.3847/2041-8213/ab14f7}

\bibitem[{{Hoang} {et~al.}(2018){Hoang}, {Naoz}, {Kocsis}, {Rasio}, \&
  {Dosopoulou}}]{hoang2018}
{Hoang}, B.-M., {Naoz}, S., {Kocsis}, B., {Rasio}, F.~A., \& {Dosopoulou}, F.
  2018, \apj, 856, 140, \dodoi{10.3847/1538-4357/aaafce}

\bibitem[{{Hopkins} {et~al.}(2006){Hopkins}, {Hernquist}, {Cox}, {Di Matteo},
  {Robertson}, \& {Springel}}]{Hopkins+06}
{Hopkins}, P.~F., {Hernquist}, L., {Cox}, T.~J., {et~al.} 2006, ApJS, 163, 1,
  \dodoi{10.1086/499298}

\bibitem[{{Inoue} {et~al.}(2020){Inoue}, {Matsushita}, {Nakanishi}, \&
  {Minezaki}}]{inoue}
{Inoue}, K.~T., {Matsushita}, S., {Nakanishi}, K., \& {Minezaki}, T. 2020, The
  Astrophysical Journal Letters, 892, L18, \dodoi{10.3847/2041-8213/ab7b7e}

\bibitem[{{Ivanov} {et~al.}(2005){Ivanov}, {Polnarev}, \& {Saha}}]{Ivanov+05}
{Ivanov}, P.~B., {Polnarev}, A.~G., \& {Saha}, P. 2005, \mnras, 358, 1361,
  \dodoi{10.1111/j.1365-2966.2005.08843.x}

\bibitem[{{Kelley} {et~al.}(2019){Kelley}, {Haiman}, {Sesana}, \&
  {Hernquist}}]{kelley}
{Kelley}, L.~Z., {Haiman}, Z., {Sesana}, A., \& {Hernquist}, L. 2019, \mnras,
  485, 1579, \dodoi{10.1093/mnras/stz150}

\bibitem[{{King}(2003)}]{king}
{King}, A. 2003, The Astrophysical Journal, 596, L27, \dodoi{10.1086/379143}

\bibitem[{{Kocsis} \& {Levin}(2012)}]{Kocsis_Levin2012}
{Kocsis}, B., \& {Levin}, J. 2012, \prd, 85, 123005,
  \dodoi{10.1103/PhysRevD.85.123005}

\bibitem[{{Kormendy} \& {Ho}(2013)}]{kormendy}
{Kormendy}, J., \& {Ho}, L.~C. 2013, Annual Review of Astronomy and
  Astrophysics, 51, 511, \dodoi{10.1146/annurev-astro-082708-101811}

\bibitem[{{Kozai}(1962)}]{kozai1962}
{Kozai}, Y. 1962, \aj, 67, 591, \dodoi{10.1086/108790}

\bibitem[{{Li} {et~al.}(2015){Li}, {Naoz}, {Kocsis}, \& {Loeb}}]{Li+15}
{Li}, G., {Naoz}, S., {Kocsis}, B., \& {Loeb}, A. 2015, \mnras, 451, 1341,
  \dodoi{10.1093/mnras/stv1031}

\bibitem[{{Lidov}(1962)}]{lidov1962}
{Lidov}, M.~L. 1962, \planss, 9, 719, \dodoi{10.1016/0032-0633(62)90129-0}

\bibitem[{{Lidov} \& {Ziglin}(1976)}]{lidov1976}
{Lidov}, M.~L., \& {Ziglin}, S.~L. 1976, Celestial Mechanics, 13, 471,
  \dodoi{10.1007/BF01229100}

\bibitem[{{Lithwick} \& {Naoz}(2011)}]{lithwick2011}
{Lithwick}, Y., \& {Naoz}, S. 2011, \apj, 742, 94,
  \dodoi{10.1088/0004-637X/742/2/94}

\bibitem[{{Liu} \& {Lai}(2020)}]{liu2020}
{Liu}, B., \& {Lai}, D. 2020, arXiv e-prints, arXiv:2004.10205.
\newblock \doarXiv{2004.10205}

\bibitem[{{Luna} {et~al.}(2019){Luna}, {Minniti}, \&
  {Alonso-Garc{\'\i}a}}]{Luna+19}
{Luna}, A., {Minniti}, D., \& {Alonso-Garc{\'\i}a}, J. 2019, \apjl, 887, L39,
  \dodoi{10.3847/2041-8213/ab5c27}

\bibitem[{{Mastrobuono-Battisti} {et~al.}(2014){Mastrobuono-Battisti},
  {Perets}, \& {Loeb}}]{MBattisti2014}
{Mastrobuono-Battisti}, A., {Perets}, H.~B., \& {Loeb}, A. 2014, \apj, 796, 40,
  \dodoi{10.1088/0004-637X/796/1/40}

\bibitem[{{McKernan} {et~al.}(2012){McKernan}, {Ford}, {Lyra}, \&
  {Perets}}]{McKernan2012}
{McKernan}, B., {Ford}, K.~E.~S., {Lyra}, W., \& {Perets}, H.~B. 2012, \mnras,
  425, 460, \dodoi{10.1111/j.1365-2966.2012.21486.x}

\bibitem[{{Meiron} {et~al.}(2017){Meiron}, {Kocsis}, \& {Loeb}}]{Meiron+2017}
{Meiron}, Y., {Kocsis}, B., \& {Loeb}, A. 2017, \apj, 834, 200,
  \dodoi{10.3847/1538-4357/834/2/200}

\bibitem[{{Meiron} \& {Laor}(2013)}]{Meiron+13}
{Meiron}, Y., \& {Laor}, A. 2013, \mnras, 433, 2502,
  \dodoi{10.1093/mnras/stt922}

\bibitem[{{Merritt}(2006)}]{Merritt2006}
{Merritt}, D. 2006, \apj, 648, 976, \dodoi{10.1086/506139}

\bibitem[{{Mezcua}(2017)}]{mezcua2017}
{Mezcua}, M. 2017, International Journal of Modern Physics D, 26, 1730021,
  \dodoi{10.1142/S021827181730021X}

\bibitem[{{Naoz}(2016)}]{naoz2016}
{Naoz}, S. 2016, Annual Review of Astronomy and Astrophysics, 54, 441,
  \dodoi{10.1146/annurev-astro-081915-023315}

\bibitem[{{Naoz} {et~al.}(2013){Naoz}, {Kocsis}, {Loeb}, \& {Yunes}}]{naoz2013}
{Naoz}, S., {Kocsis}, B., {Loeb}, A., \& {Yunes}, N. 2013, \apj, 773, 187,
  \dodoi{10.1088/0004-637X/773/2/187}

\bibitem[{{Naoz} \& {Silk}(2014)}]{NaozSilk14}
{Naoz}, S., \& {Silk}, J. 2014, \apj, 795, 102,
  \dodoi{10.1088/0004-637X/795/2/102}

\bibitem[{{Naoz} {et~al.}(2019){Naoz}, {Silk}, \& {Schnittman}}]{Naoz+19}
{Naoz}, S., {Silk}, J., \& {Schnittman}, J.~D. 2019, \apjl, 885, L35,
  \dodoi{10.3847/2041-8213/ab4fed}

\bibitem[{{Naoz} {et~al.}(2020){Naoz}, {Will}, {Ramirez-Ruiz}, {Hees}, {Ghez},
  \& {Do}}]{Naoz+20}
{Naoz}, S., {Will}, C.~M., {Ramirez-Ruiz}, E., {et~al.} 2020, \apjl, 888, L8,
  \dodoi{10.3847/2041-8213/ab5e3b}

\bibitem[{{Neumayer} {et~al.}(2020){Neumayer}, {Seth}, \&
  {B{\"o}ker}}]{neumayer}
{Neumayer}, N., {Seth}, A., \& {B{\"o}ker}, T. 2020, \aapr, 28, 4,
  \dodoi{10.1007/s00159-020-00125-0}

\bibitem[{{O'Leary} {et~al.}(2009){O'Leary}, {Kocsis}, \& {Loeb}}]{oleary}
{O'Leary}, R.~M., {Kocsis}, B., \& {Loeb}, A. 2009, \mnras, 395, 2127,
  \dodoi{10.1111/j.1365-2966.2009.14653.x}

\bibitem[{{Peters}(1964)}]{Peters1964}
{Peters}, P.~C. 1964, Physical Review, 136, 1224,
  \dodoi{10.1103/PhysRev.136.B1224}

\bibitem[{{Portegies Zwart} {et~al.}(2006){Portegies Zwart}, {Baumgardt},
  {McMillan}, {Makino}, {Hut}, \& {Ebisuzaki}}]{pzwart2006}
{Portegies Zwart}, S.~F., {Baumgardt}, H., {McMillan}, S.~L.~W., {et~al.} 2006,
  \apj, 641, 319, \dodoi{10.1086/500361}

\bibitem[{{Randall} \& {Xianyu}(2019)}]{randall2019}
{Randall}, L., \& {Xianyu}, Z.-Z. 2019, arXiv e-prints, arXiv:1902.08604.
\newblock \doarXiv{1902.08604}

\bibitem[{{Rasskazov} {et~al.}(2019){Rasskazov}, {Fragione}, {Leigh}, {Tagawa},
  {Sesana}, {Price-Whelan}, \& {Rossi}}]{Rasskazov+19}
{Rasskazov}, A., {Fragione}, G., {Leigh}, N. W.~C., {et~al.} 2019, \apj, 878,
  17, \dodoi{10.3847/1538-4357/ab1c5d}

\bibitem[{{Robertson} {et~al.}(2006){Robertson}, {Bullock}, {Cox}, {Di Matteo},
  {Hernquist}, {Springel}, \& {Yoshida}}]{Robertson+06}
{Robertson}, B., {Bullock}, J.~S., {Cox}, T.~J., {et~al.} 2006, \apj, 645, 986,
  \dodoi{10.1086/504412}

\bibitem[{{Sesana} {et~al.}(2011){Sesana}, {Gualandris}, \&
  {Dotti}}]{Sesana+11}
{Sesana}, A., {Gualandris}, A., \& {Dotti}, M. 2011, \mnras, 415, L35,
  \dodoi{10.1111/j.1745-3933.2011.01073.x}

\bibitem[{{Tagawa} {et~al.}(2019){Tagawa}, {Haiman}, \& {Kocsis}}]{tagawa2019}
{Tagawa}, H., {Haiman}, Z., \& {Kocsis}, B. 2019, arXiv e-prints,
  arXiv:1912.08218.
\newblock \doarXiv{1912.08218}

\bibitem[{{Wegg} \& {Nate Bode}(2011)}]{Wegg+11}
{Wegg}, C., \& {Nate Bode}, J. 2011, \apjl, 738, L8,
  \dodoi{10.1088/2041-8205/738/1/L8}

\bibitem[{{Yu} \& {Tremaine}(2003)}]{Yu+03}
{Yu}, Q., \& {Tremaine}, S. 2003, \apj, 599, 1129, \dodoi{10.1086/379546}

\bibitem[{{Yunes} {et~al.}(2011){Yunes}, {Miller}, \&
  {Thornburg}}]{Yunes_Miller2011}
{Yunes}, N., {Miller}, M.~C., \& {Thornburg}, J. 2011, \prd, 83, 044030,
  \dodoi{10.1103/PhysRevD.83.044030}

\end{thebibliography}
\bibliographystyle{aasjournal}

\end{document}